\documentclass[preprint,12pt]{elsarticle}
\usepackage{amssymb}
\usepackage{lineno}
\usepackage{xcolor}
\usepackage[export]{adjustbox}
\usepackage{float}
\usepackage{booktabs}
\usepackage{multirow}
\usepackage{siunitx}
\usepackage{textcomp}
\usepackage{longtable}
\usepackage{amsmath}
\usepackage{systeme}
\usepackage[normalem]{ulem} 

\usepackage[labelformat=simple]{subcaption}

\newsavebox{\leftpic}
\newsavebox{\rightpic}

\usepackage{hyperref}
\hypersetup{
    colorlinks=true,
    linkcolor=blue,
    filecolor=magenta,      
    urlcolor=blue,
}

\begin{document}

\begin{frontmatter}

\title{
Film-wise condensation of pure vapour in flattened tubes: A numerical study of the combined influence of aspect ratio and rotation angle}

\author[1]{Y. V. Lyulin}

\author[2]{D. Bugrov}

\author[2]{R. Khurmatova}

\author[2]{H. Ouerdane}

\author[3,4]{I. Marchuk}

\address[1]{National Research University ``Moscow Power Engineering Institute'', Krasnokazarmennaya 17, Moscow 111250, Russia}

\address[2]{Center for Digital Engineering, Skolkovo Institute of Science and Technology, 30 Bolshoy Boulevard, Moscow 121205, Russia
}

\address[3]{Novosibirsk State University, 2 Pirogov Street, Novosibirsk, 630090, Russia}

\address[4]{Kutateladze Institute of Thermophysics SB RAS, 1, Ac. Lavrentieva ave., Novosibirsk, 630090, Russia}

\begin{abstract}
Vapor condensation is a physical phenomenon that finds application in heat removal systems. The traditional design of these systems involves round tubes but experience shows that this geometry is not optimal for heat transfer. Flattened tubes on the other hand, have been found to offer potential for improvement as their geometry increases the condensation surface, which fosters higher heat transfer rates. However, the effects of tube shape (aspect ratio) and orientation (rotation angle) on film-wise condensation dynamics are not fully understood. In this work, we numerically simulate a model of the condensed vapor layer thickness distribution on the flattened tube inner surfaces taking into account bulk and surface forces (gravity, surface tension, shear stress) for a thin layer of liquid. We consider various configurations of aspect ratios (circular, and AR = 2, 4, and 6) and rotation angles (0$^{\circ}$, 10$^{\circ}$, 20$^{\circ}$, 30$^{\circ}$, 45$^{\circ}$, 60$^{\circ}$, 75$^{\circ}$, and 90$^{\circ}$). Our simulations allow for an improved understanding of how these geometric parameters as well as their interplay, influence the thickness distribution of the condensate film on the tube's inner surface, and facilitate the identification of configurations that maximize heat transfer efficiency. Considering water as a working fluid, results show a possible heat transfer enhancement of up to 74\% compared to the round tube geometry for an aspect ratio of 6 and a rotation angle of 90$^{\circ}$.
\end{abstract}


\end{frontmatter}

\noindent
    \parbox{\linewidth}{%
    Highlights
    \begin{itemize}
        \item Both capillary pressure and gravity control the liquid film's shape in the tube;
        \item The liquid accumulates in the tube's rounded areas; flat parts remain covered with thin films;
        \item The rotation angle influences the condensate's distribution inside the tube;
        \item A maximum heat transfer gain of 74\% is achieved at AR 6 and 90$^{\circ}$ rotation angle.
    \end{itemize}
}

\section*{Nomenclature}

\noindent
\textbf{Mathematical symbols}\\
\begin{tabular}{ll}
$D_{\rm h}$     &hydraulic diameter \\
$G$             & mass flow [kg$\cdot$m$^{-2}\cdot$s$^{-1}$] \\
$L$             & distance along the surface [mm] \\
$R$             & radius \\
$S$             & surface \\
$T$             & temperature [$^{\circ}$C] \\
$T_{\rm sat}$   & saturation temperature [$^{\circ}$C] \\
$g_{i,j}$       & metric tensor matrix of the surface \\
$p_0$           & initial pressure inside the tube [N$\cdot$m$^{-2}$] \\
$s$             & length of the curved surface along the perimeter \\
$t$             & time [s] \\
$u$             & velocity of liquid [m$\cdot$s$^{-1}$] \\
$V$             & liquid volume [m$^{3}$] \\
$h$             & liquid thickness [mm] \\
$h_f$           & film thickness [$\mu$m] \\
$n$             & unit vector of the surface \\
$x$             & vapor quality \\
%
$\Delta{t}$         & time step [s] \\
$\Psi$              & latent heat of vaporization \\
$\xi, \zeta, \eta$  & coordinate system of the surface \\
$\theta$            & angle of curve rotation [rad] \\
$\rho$              & density [kg$\cdot$m$^{-3}$] \\
$\lambda$           & thermal conductivity [W$\cdot$m$^{-1}\cdot~^{\circ}$C$^{-1}$] \\
$\mu$               & viscosity [N$\cdot$s$\cdot$m$^{-2}$] \\
$\nu$               & kinematic viscosity [N$\cdot$s$\cdot$m$\cdot$kg$^{-1}$] \\
$\sigma$            & surface tension [N$\cdot$m$^{-1}$] \\
$\tau$              & tangential unit vector \\
$\phi$              & void fraction \\
$\Omega$            & element of the substrate surface \\
\end{tabular}
\\~\\

\textbf{Subscripts}\\
\begin{tabular}{ll}
$w$ & water \\
$m$ & mass \\
$v$ & vapor \\
\end{tabular}
\\

\textbf{Abbreviations}\\
\begin{tabular}{ll}
AR  &the aspect ratio of flattened tubes \\
FT  &flattened tube \\
RT  &rounded tube \\
ST  &square tube \\
HTC &heat transfer coefficient \\
EF  &enhancement factor \\
\end{tabular}


\section{Introduction}
\label{sec1}
The study of film-wise vapor condensation inside mini-channels for heat transfer \cite{ElKadi2021} is of great importance for many applications such as, e.g., air conditioning systems \cite{Han2012}, compact heat exchangers \cite{Li2011}, and cooling devices for micro and power electronics \cite{MANOVA2020114669,Zhang2021}, as the latent heat of vaporization of the working fluid is used to efficiently transfer heat at a nearly constant temperature. Mini-channels in thermostabilization systems provide several advantages, including compact size, high heat transfer coefficients, reduced cost and a minimized volume of refrigerants. These characteristics are particularly critical in the aerospace sector where stringent requirements on mass, volume, and reliability necessitate highly efficient thermal management solutions. Systems such as heat pipes, loop heat pipes, capillary pumped loops, and mechanically pumped loops, which rely on phase-change mechanisms to transfer heat from the core module equipment to external radiators are widely implemented in satellites \cite{marchuk2013theoretical, lyulin2015investigation}. However, a better understanding of film-wise vapor condensation in mini-channels remains key to optimize heat exchangers, improve their performance and extend their range of applicability.
    
Different passive approaches are used to enhance heat transfer efficiency in condensers. Techniques such as surface roughening, using micro-fins, or integrating porous materials can significantly improve thermal performance by increasing the surface area and raising turbulence in flow \cite{belyaev2021study}. However, these enhancements come with a trade-off like an increase of pressure drop, which precludes optimal performance. Thus, it is crucial to maintain a balance between enhanced thermal performance and acceptable pressure drops to ensure an overall cooling system efficiency. Achieving this balance involves careful design and optimization of the condenser's geometry and materials, taking into account the specific operational requirements and system limitations. 

An effective method to improve vapor condensation is the use of curvilinear surfaces and finned tubes \cite{glushuk2017experemental, marchuk2006vapor}. When applying this method of heat transfer enhancement, together with an increase of the thermal exchange surface area, the mean-integral film thickness is significantly reduced due to the action of capillary pressure \cite{marchuk2003problem, marchuk2016model, miscevic2007condensation, serin2009miniaturised}. Another solution to increase the condensation heat transfer is by using mini- and micro-channels with non-circular geometry, including flattened tubes, which can be obtained by modifying round tubes. The transition to the flattened tubes changes the development of thermal and hydraulic boundary layers formation on the tube surface in a way that enhances heat transfer. Effects of tube shape on condensation heat transfer are discussed in Del Col et al. \cite{delcol2011effect}, Liu et al. \cite{liu2016experimental}, Derby et al. \cite{derby2012condensation}, Agarwal et al. \cite{agarwal2010measurement}. Mghari et al. \cite{elmghari2014condensation}. Bortolin et al. \cite{bortolin2014condensation}, Wang and Rose \cite{wang2005filmcondensation}, and Gu et al. \cite{xingu2019effect}. Further, the inclination of the condenser tube also significantly influences heat transfer performance \cite{stephane2011twophase, william2019effect, lyulin2011experimental, cavallini2003condensation}. Lips and Meyer \cite{stephane2011twophase} provide an overview of two-phase flows in inclined tubes, with specific reference to condensation. Notably, a crucial factor in two-phase flows is the void fraction, and accurately predicting it is a significant challenge in applications involving two-phase flows. 

Experimental results reported in the works of William et al. \cite{william2019effect} and Lyulin et al. \cite{lyulin2011experimental} clearly show an increase in the average overall heat transfer coefficient for downwardly inclined tubes with respect to the horizontal direction. The maximum heat transfer coefficient during vapor condensation in round smooth tubes is achieved for downward tube inclination between \ang{15} and \ang{35} \cite{lyulin2011experimental}. This increase is primarily due to improved drainage of condensate with increasing inclination angle. More experimental research that would include the measurement of void fraction is needed for a better understanding of the impact of inclination angle. 

The present work is focused on the numerical modeling of the condensation process of a pure vapor inside a flattened tube considering various aspect ratios and rotation angles. We analyze the effect of pipe rotation on the film condensate distribution on the inner surface of the tube and on heat transfer. We find that capillary pressure and gravity together control the film's shape, with liquid accumulation in the rounded areas while the flat parts remain covered with a thin film. We also study how rotation of the pipe influences the condensate distribution. The maximum heat transfer gain, up to 74\% is achieved for an aspect ratio of 6 and a rotation angle of  90$^\circ$.
    
The paper is organized as follows. Section~\ref{sec2} is devoted to a literature review focused on two-phase flows and heat transfer. A theoretical model for film-wise condensation of pure vapour is presented in Section~\ref{sec3}. The calculation results are discussed in Section \ref{sec4}. Section \ref{sec5} concludes the paper.

\section{Literature review}
\label{sec2}
A significant number of works on two-phase flow and heat transfer characteristics at the vapor condensation are devoted to the study of round and horizontal tubes. The reviews by Cavallini et al. \cite{cavallini2003condensation} and Dalkilic and Wongwises \cite{dalkilic2013condensation} offer a comprehensive analysis of condensation phenomena within and around smooth and enhanced tubes, highlighting advancements in heat transfer mechanisms. Their work synthesizes experimental and theoretical findings, providing a foundation to optimizing condensation processes for industrial applications. An insightful review on condensation heat transfer in micro-channels and mini-channels is given in Awad et al. \cite{awad2014critical}. An experimental study of humidified air condensation in a serpentine-type heat exchanger was conducted by Po\v{s}kas et al. \cite{povskas2024experimental}, focusing on the effect of cooling water flow rate on condensation across different tube rows. The study covered a wide range of conditions, including different Reynolds numbers, water vapor mass fractions, and inlet air temperatures. The results show that the cooling ratio has a significant influence on the local condensation flux and efficiency. In particular, an optimal cooling ratio of 3 was found to maximize performance. In a study by Shah \cite{shah2019prediction}, several correlations for predicting condensation heat transfer in non-circular channels were evaluated against extensive experimental data. Discrepancies were identified in existing models and the author emphasized the need for more accurate predictive tools, particularly for non-standard geometries in heat exchangers. The experimental data encompassed a wide range of conditions, including varying channel shapes, refrigerant types, and flow regimes. The author concluded that more studies of the condensation of pure vapor inside flattened tubes are needed.
    
Condensation of pure vapor in flattened tubes has been the subject of numerous theoretical \cite{stefano2010numerical, stefano2012numerical, zhang2016numerical, weili2017numerical, jian2018numerical, li2020condensation, alnaimat2023cfd} and experimental studies \cite{wilson2003refrigerant, kim2013condensation, lee2014condensation, darzi2015experimental, kaewon2016condensation, ghorbani2017experimental, anand2019condensation, sereda2021heat, rukruang2023experimental, cheng2024heat}. These works were mainly devoted to studying the effect of the aspect ratio on the condensation process and comparing the results concerning round pipes. A summary of technical details of the systems studied in these articles is given in Table \ref{table:1}.

\begin{table}[ht!]
    \resizebox{\textwidth}{!}{%
    \begin{tabular}{|l|l|l|l|l|l|}
    \hline
    Author &
    Fluid &
    Geometry &
    G, kg$\cdot$m$^{-2}\cdot$s$^{-1}$ & $T_{\rm sat}$, $^\circ$C &
    \begin{tabular}[c]{@{}l@{}}Vapour \\ quality\end{tabular} \\ \hline\hline
    
    \multicolumn{6}{|l|}{{\bf Theory and simulation}} \\ \hline\hline
    
    \begin{tabular}[c]{@{}l@{}} Nebuloni and \\ Thome (2010) \cite{stefano2010numerical} \end{tabular} &
    
    \begin{tabular}[c]{@{}l@{}} R134a \\ Ammonia\end{tabular} &
    \begin{tabular}[c]{@{}l@{}} FT: AR = 2 \\ ($D_{\rm h}$ = 0.54 - 1.1 mm)\end{tabular} &
    300 &
    \begin{tabular}[c]{@{}l@{}} 30 \\ 10 \end{tabular} &
    0.98 \\ \hline
      
    \begin{tabular}[c]{@{}l@{}}Nebuloni and \\ Thome (2012) \cite{stefano2012numerical} \end{tabular} &
      R134a &
      FT: AR = 2 ($D_{\rm h}$ = 0.27 mm) &
      300 &
      40 &
      0.99 \\ \hline
      
    \begin{tabular}[c]{@{}l@{}}Zhang et al. \\ (2016) \cite{zhang2016numerical}\end{tabular} &
      \begin{tabular}[c]{@{}l@{}} R410A \\ R134a \end{tabular} &
      \begin{tabular}[c]{@{}l@{}}RT: $D_{\rm h}$ = 3.78 mm\\ FT: AR = 3.07, 4.23, 5.29 \end{tabular} &
      421 - 1083 &
      47 &
      0.4 - 0.95 \\ \hline
      
    \begin{tabular}[c]{@{}l@{}} Li et al. \\ (2017) \cite{weili2017numerical}\end{tabular} &
      R410A &
      \begin{tabular}[c]{@{}l@{}} RT: $D_{\rm h}$ = 3.78 mm \\ FT: AR = 3.07, 4.23, 5.39 \end{tabular} &
      305 - 1061 &
      47 &
      0.41 - 0.98 \\ \hline
      
    \begin{tabular}[c]{@{}l@{}} Wen et al. \\ (2018) \cite{jian2018numerical} \end{tabular} &
      R134a &
      \begin{tabular}[c]{@{}l@{}} RT: $D_{\rm h}$ = 3.25, 4.57, 6.5 mm \\ FT: AR = 2, 4, 6 \end{tabular} &
      600 - 1000 &
      40 &
      0.4 - 0.95 \\ \hline
    
    \begin{tabular}[c]{@{}l@{}} Wei Li, Di Lyu \\ (2020) \cite{li2020condensation}\end{tabular} &
      R134a &
      FT: AR = 0.4, 0.8 &
      200 - 400 &
      60 &
      0.5 - 0.8 \\ \hline

    \begin{tabular}[c]{@{}l@{}} Alnaimat et al. \\ (2023) \cite{alnaimat2023cfd} 
    \end{tabular} &
      \begin{tabular}[c]{@{}l@{}} R134a \\ Propane (R290) \end{tabular} &
      ST 0.5 mm &
      150 - 1200 &
      40 &
      0.5 - 1 \\ \hline\hline
      
    \multicolumn{6}{|l|}{{\bf Experiments}} \\ \hline\hline
    
    \begin{tabular}[c]{@{}l@{}}Wilson et al. \\ (2003) \cite{wilson2003refrigerant} \end{tabular} &
      \begin{tabular}[c]{@{}l@{}} R410A \\ R134a \end{tabular} &
      \begin{tabular}[c]{@{}l@{}} RT: $D_{\rm h}$ = 8.91 mm \\ FT: AR = 0.87, 1.8, 3.87, 12.83 \end{tabular} &
      75 - 400 &
      35 &
      0.1 - 0.8 \\ \hline
      
    \begin{tabular}[c]{@{}l@{}} Kim et al. \\ (2013) \cite{kim2013condensation}\end{tabular} &
      R410A &
      \begin{tabular}[c]{@{}l@{}} RT: $D_{\rm h}$ = 5 mm \\ FT: AR = 2, 4, 6 \end{tabular} &
      100 - 400 &
      45 &
      0.2 - 0.8 \\ \hline
      
    \begin{tabular}[c]{@{}l@{}} Lee et al. \\ (2014) \cite{lee2014condensation}\end{tabular} &
      R410A &
      \begin{tabular}[c]{@{}l@{}} RT: $D_{\rm h}$ = 7 mm \\ FT: AR = 2, 4 \end{tabular} &
      100 - 400 &
      45 &
      0.2 - 0.8 \\ \hline
      
    \begin{tabular}[c]{@{}l@{}}Darzi et al. \\ (2015) \cite{darzi2015experimental}\end{tabular} &
      R600a &
      \begin{tabular}[c]{@{}l@{}} RT: $D_{\rm h}$ = 8.7 mm \\ FT: AR = 0.46, 1.05, 2.84 \end{tabular} &
      154.8 - 265.4 &
      47 &
      0.1 - 0.8 \\ \hline
      
    \begin{tabular}[c]{@{}l@{}}Kaew-On et al. \\ (2016) \cite{kaewon2016condensation} \end{tabular} &
      R134a &
      \begin{tabular}[c]{@{}l@{}} RT: $D_{\rm h}$ = 3.51 mm \\ FT: AR = 0.72, 3.49, 7.02 \end{tabular} &
      350 - 900 &
      31.3 - 46.3 &
      0.1 - 0.9 \\ \hline
      
    \begin{tabular}[c]{@{}l@{}} Ghorbani et al. \\ (2018) \cite{ghorbani2017experimental} \end{tabular} &
      \begin{tabular}[c]{@{}l@{}} R600a \\ R600a-oil \\ R600a-oil-CuO \end{tabular} &
      FT: AR = 2 &
      110 - 372 &
      36.2 - 45.6 &
      0.1 - 0.85 \\ \hline
      
    \begin{tabular}[c]{@{}l@{}} Solanki and \\ Kumar (2019) \cite{anand2019condensation} \end{tabular} &
      R-134a &
      \begin{tabular}[c]{@{}l@{}} RT: $D_{\rm h}$ = 8.91 mm \\ FT: AR = 2.72, 5.8 \end{tabular} &
      450 - 650 &
      35 - 45 &
      0.1 - 0.8 \\ \hline
    
    \begin{tabular}[c]{@{}l@{}} Sereda and Rifert. \\  (2021) \cite{sereda2021heat} \end{tabular} &
      R22, R407 &
      RT: $D_{\rm h}$ = 17 mm  &
      57 &
      40 &
      0.23 - 0.95 \\ \hline
    
    \begin{tabular}[c]{@{}l@{}} Rukruang. \\  (2023) \cite{rukruang2023experimental} \end{tabular} &
      R32 &
      \begin{tabular}[c]{@{}l@{}} RT: $D_{\rm h}$ = 4.31 mm \\ \end{tabular} &
      280 - 580 &
      40 - 50 &
      0.1 - 0.8 \\ \hline

    \begin{tabular}[c]{@{}l@{}} Cheng et al. \\ (2024) \cite{cheng2024heat}
    \end{tabular} &
      \begin{tabular}[c]{@{}l@{}} Methanol \\ Water \end{tabular} &
      RT: $D_{\rm h}$ = 5.6 mm &
      \begin{tabular}[c]{@{}l@{}} 990 \\ 1250 \end{tabular} &
      \begin{tabular}[c]{@{}l@{}} 40 - 100 \\ 40 - 90 \end{tabular} &
      0.35 - 0.65 \\ \hline

    \end{tabular}%
    }
    \caption{Works devoted to condensation in flattened tubes with different working fluids, mass flow rates, saturation temperatures and geometries -- RT: round tubes; FT: flattened tube; AR: aspect ratio.}
    \label{table:1}
\end{table}

Nebuloni and Thome \cite{stefano2010numerical, stefano2012numerical} numerically investigated annular laminar film condensation in flattened microtubes focusing on film profile and liquid film thickness. The annular condensation heat transfer coefficients were found to be strongly dependent on the channel geometrical shapes and on the distribution of heat fluxes. Numerical simulations of the condensation heat transfer coefficients and pressure gradients of R410A and R134a in horizontal round and flattened tubes were reported in the work of Zhang et al. \cite{zhang2016numerical}. The local heat transfer coefficients and pressure gradients were found to increase with the rise of the mass flux, the vapor quality, and the aspect ratio. Heat transfer characteristics for the condensation of R410A inside an horizontal round tube (with a diameter of 3.78 mm) and a flattened tube (with AR = 3.07, 4.23, and 5.39) with larger horizontal dimensions than vertical at a saturation temperature of 320 K were numerically investigated by Li et al. \cite{weili2017numerical}. They obtained that heat transfer coefficients of the flattened tubes were about 1.5 times larger than they are with round tubes with mass flow $G = 1061$ kg$\cdot$m$^{-2} \cdot $s$^{-1}$, and vapor quality $x = 0.8$. The liquid film in the flattened tube accumulates on the sides of the bottom surface and in the middle of the top surface of the channels when the vapor quality is low. Wen et al. \cite{jian2018numerical} numerically studied the condensation performance of R134a in horizontal round tubes and flattened tubes. Three round tubes with different hydraulic diameters ($D_{\rm h}$ = 3.25, 4.57 and 6.5 mm) were deformed into flattened tubes with the AR equal to 2, 4 and 6, respectively. The liquid film thickness in combination with the interface line was shown in detail to illustrate the effects of shear stress, surface tension and gravity. The heat transfer coefficients increase with the vapor quality and aspect ratio, and the enhancement is magnified at higher vapor quality and mass velocity. Three-dimensional simulations were carried out at different mass fluxes, vapor qualities, and gravity conditions in the study by Li et al. \cite{li2020condensation}. Simulation results reveal that the liquid film tends to accumulate at the corners of the cross-section of the mini-channel. Furthermore, the thickness of the liquid film increases as mass flux decreases and vapor quality decreases. In the CFD study by Alnaimat et al. \cite{alnaimat2023cfd} on condensation heat transfer and flow regime characteristics of R134a and propane in a square microchannel, results show that propane provides an enhanced heat transfer coefficient by 65–80 \% compared to R134a while maintaining efficient condensation behavior.
    
Changes in the flow field characteristics of the refrigerants R134a and R410A were experimentally studied by Wilson et al. \cite{wilson2003refrigerant}. Horizontal smooth and micro-finned round tubes of 9 mm were deformed into a flattened tubes, which showed a significant reduction in refrigerant charge. The heat transfer coefficients increase with the aspect ratio and vapor quality for the earlier transition from stratified flow to annular flow, as well as for different flow field configurations formed in round and flattened tubes. Kim et al.\cite{kim2013condensation} reported experimental data on the local condensation heat transfer coefficients and pressure gradients at the condensation of R410a vapor in a horizontal stainless round tube of 5 mm and flattened tubes with various aspect ratios of 2, 4, and 6. In the annular flow regime, an enhancement of the heat transfer coefficient was observed with the increase of the aspect ratio. Condensation heat transfer coefficients and pressure drops of R-410A in flattened micro fin tubes made from 7.0 mm round micro fin tubes have been reported \cite{lee2014condensation}. Results show that the effect of the aspect ratio on the condensation heat transfer coefficient is dependent on the flow pattern. For an annular flow, the heat transfer coefficient increases as the aspect ratio increases. However, in the stratified flow regime the heat transfer coefficient decreases and the pressure drop increases as the aspect ratio increases. A possible explanation can be based on the estimated flow pattern in flat micro-fin tubes. Darzi et al. \cite{darzi2015experimental} experimentally studied the condensation of R600a in a horizontal round tube of 8.7 mm and three flattened tubes with hydraulic diameters $D_{\rm h}$ = 8.2, 7.29 and 5.1 mm. The study demonstrated the enhancement in the heat transfer coefficient and pressure drop for all the flattened tubes. This effect was more obvious for smaller hydraulic diameters. Kaew-On et al. \cite{kaewon2016condensation} experimentally investigated the condensation heat transfer characteristic of R134a flowing in a round tube and three flattened copper tubes. The following ranges: 350 - 900 kg$\cdot$m$^{-2}\cdot$s$^{-1}$ for the mass flow rate, 10 - 50 kW$\cdot$m$^{-2}$ for the heat flux, and 0.1-0.9 for the inlet quality, were covered. They found that the heat transfer coefficients of the flattened tubes were larger than that of the circular tube, from 5\% up to 400\% for various aspect ratios. An experimental investigation of condensation heat transfer of R600a/POE/CuO nano refrigerant in flattened tubes by Ghorbani et al. \cite{ghorbani2017experimental}, showed that the addition of copper oxide nanoparticles to the base refrigerant, R600a, which has polyol ester (POE) as a lubricant, improves the condensation heat transfer coefficient compared to using only the conventional refrigerant. In Solanki and Kumar study \cite{anand2019condensation}, the heat transfer coefficients and pressure drops of R-134a fluid inside round and flat tubes were investigated experimentally with mass flow rates of 450, 550, and 650 kg$\cdot$m$^{-2}\cdot$s$^{-1}$ at saturation temperatures of 35, 40, and 45$^\circ$C. The maximum enhancement factor of condensation heat transfer was achieved in a flattened tube with an aspect ratio of 5.88 at low mass flow rate (450 kg$\cdot$m$^{-2}\cdot$s$^{-1}$) and low vapor quality ($x$ = 0.1). Also, the maximum pressure drop penalty was also obtained, at high mass flux (650 kg$\cdot$m$^{-2}\cdot$s$^{-1}$) and high vapor quality ($x$ = 0.8) for the same flattened tube. Heat transfer during film condensation inside horizontal tubes in stratified phase flow was studied by Sereda et al. \cite{sereda2021heat} using a CFD model, which was experimentally validated. The obtained results allowed for improving the prediction of effective heat transfer coefficients for vapor condensation, taking into account the influence of the condensate flow in the bottom part of the tube on the heat transfer. In the experimental study by Rukruang et al. \cite{rukruang2023experimental} on condensation heat transfer and pressure drop characteristics of the R32 fluid flowing inside an alternating cross-section flattened tube, the results indicate that the flattened tube reveals a large condensation heat transfer coefficient and a reasonable pressure loss penalty compared to circular tubes. In the experimental study by Cheng et al. \cite{cheng2024heat} on a 3D fin-and-tube pulsating heat pipe heat exchanger under vertical and horizontal orientations, results reveal that methanol achieves stable pulsation-like behavior even in large-diameter tubes, enhancing heat transfer effectiveness by up to 50\% compared to systems without cross-layer flow.
     
\section{Parametrization of flattened tubes for numerical modeling}
\label{sec3}
\begin{figure}
        \centering
        \includegraphics[scale=0.25]{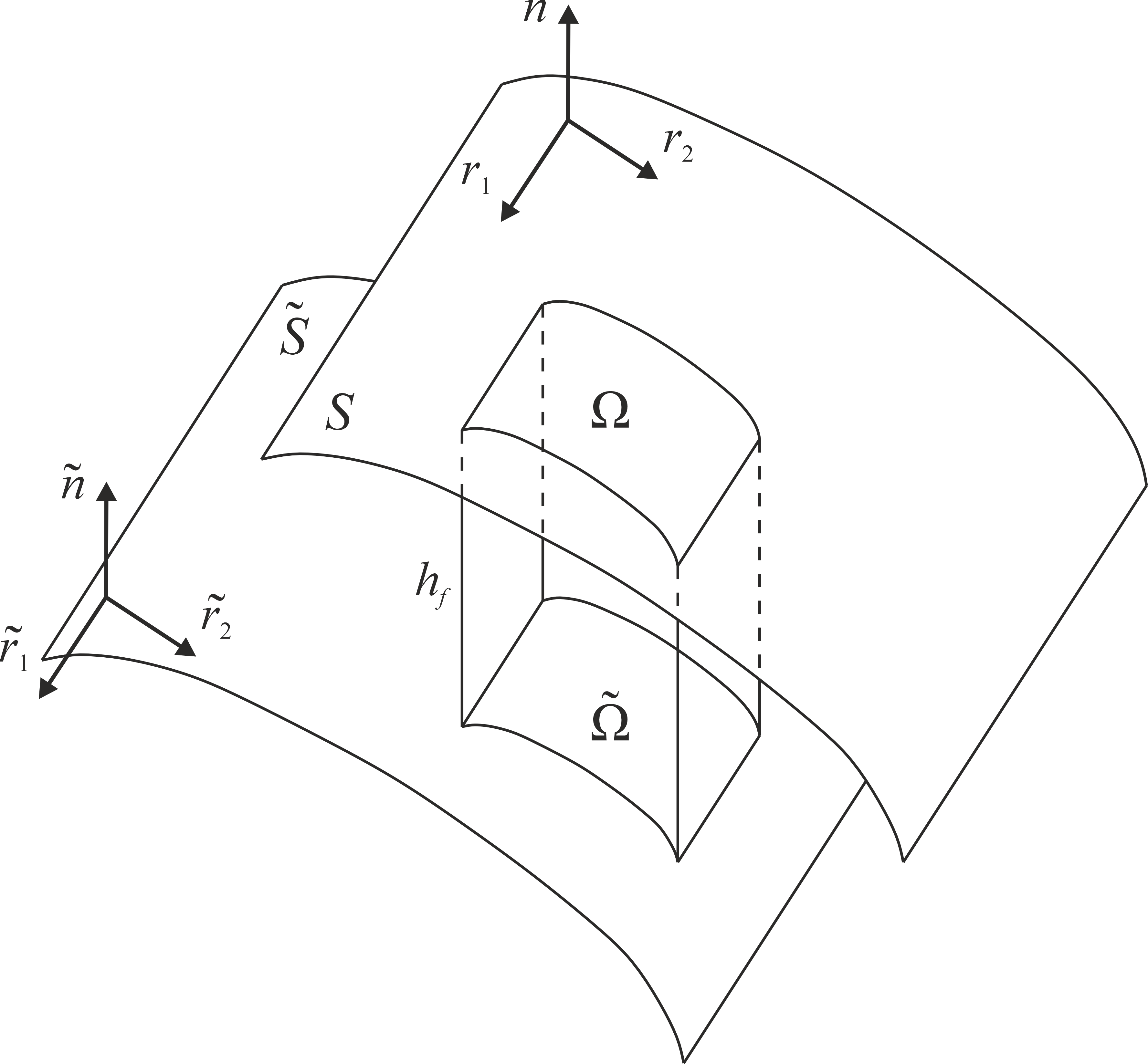}
        \caption{Condensation surface and coordinate systems.}
        \label{fig:surface}
    \end{figure}

For our numerical study of film-wise condensation of pure vapor on a curvilinear surface, we consider that a thin layer of a working substance, which is assumed to be a viscous incompressible liquid, flows on the inner surface of a tube as depicted in Fig. \ref{fig:surface}. The thickness of the film at time $t$ is given by the distance $h_f$ between its upper and lower surfaces, $S$ and $\tilde{S}$, the former being in contact with the tube inner surface. In the liquid layer, the point position is determined by the coordinates $(\xi,\zeta,\eta)$, where $(\xi,\zeta)$ are the coordinates on the surface and $\eta$ is the distance from the surface along the direction normal to the surface $S$. The vector $\tilde{r}(t, \xi, \zeta) = r + hn$ is the parametrization of the liquid's free surface $\tilde{\Omega}$. The unit vector $\tilde{n}$ is perpendicular to the surface of the liquid. 

Let $h(t,\xi,\zeta)$ be the liquid layer thickness at the moment $t$. The basic equation that we aim to solve reads \cite{Nusselt1916}:

     \begin{equation}
     \label{nonlinear}
        \begin{aligned}
             h_f + \text{div}_{S}\left[-\frac{h^3}{3\mu} \text{grad}_\text{S} (p_0 + {\rho} g \tilde{r} + \sigma\tilde{H}) + \frac{h^2}{2\mu} \tau_{sur}\right] - \\ - \frac{\lambda(T_S - T_W)}{\rho \Psi h} \sqrt{\det{\tilde{g}_{ij}}} = 0
         \end{aligned}
     \end{equation}

\noindent This equation, which generalizes models of film-wise vapor condensation, and its derivation have been discussed in Ref.~\cite{marchuk2015filmwise}. Note that for flat surfaces, similar equations describing the motion of thin liquid films are given in \cite{marchuk2016model, miscevic2007condensation}. To numerically solve Eq.~\eqref{nonlinear}, we parametrize the surface of the flattened tube (both flat and round sections). The shape of the tube can be determined by a known function of the section curvature $k(s)$, where $s$ is the length of the curve. We use the following parametrization for the coordinates, where $\alpha$ is the angle of the curve that defines the portion of the pipe:

    \begin{equation}
        \alpha (s) = \int_0^s k(\xi)d\xi
    \end{equation}

    \begin{equation}
        x (s) = \int_0^s \cos(\alpha(\xi))d\xi
    \end{equation}

    \begin{equation}
        y (s) = -\int_0^s \sin(\alpha(\xi))d\xi
    \end{equation}

    \begin{figure}[hbt!]
        \centering
        \includegraphics[width=0.7\linewidth]{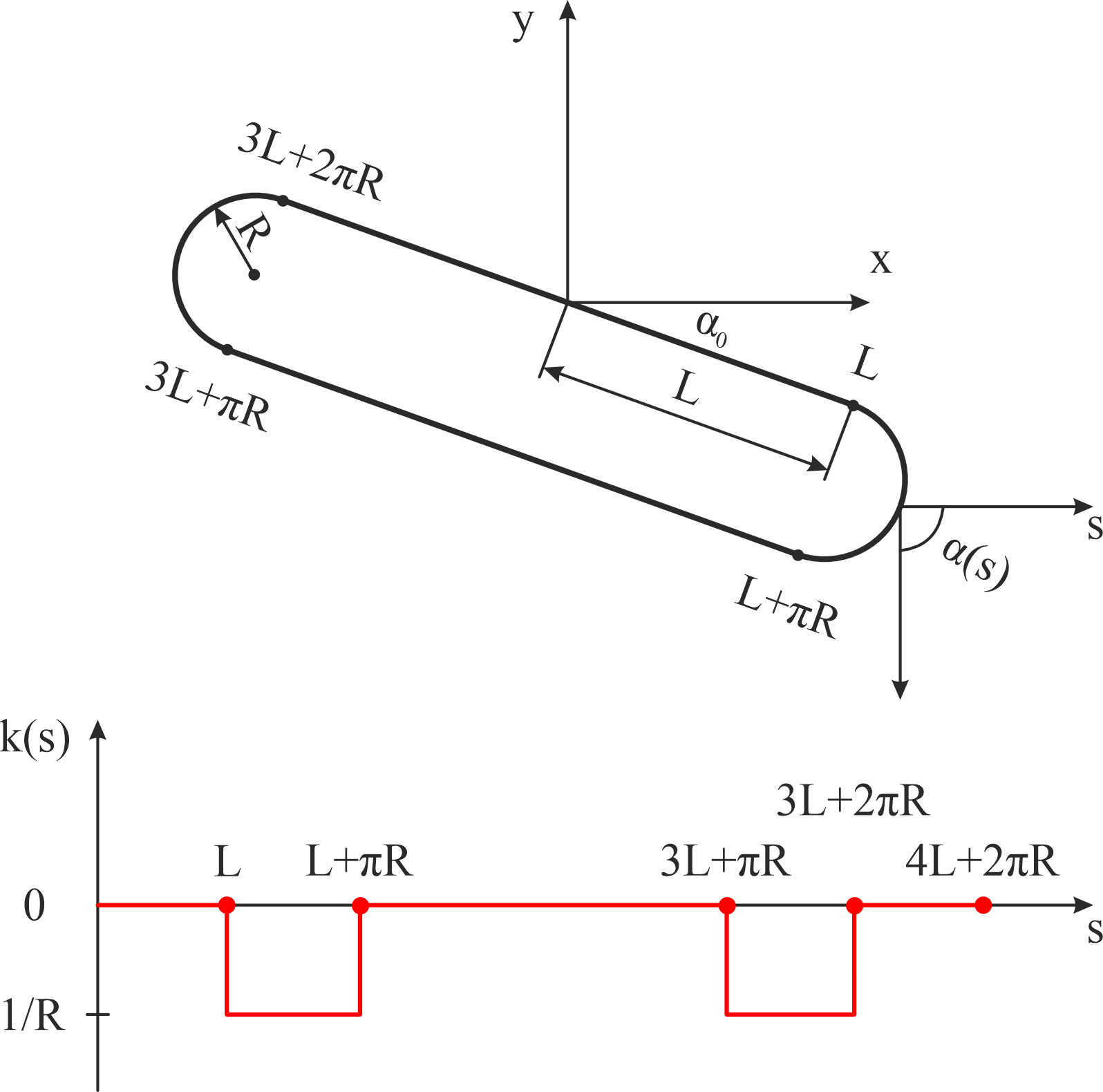}
        \caption{Section of flattened tube.}
        \label{fig:section}
    \end{figure}

To avoid numerical issues caused by sudden changes from flat to rounded shape, the curvature function is represented using the hyperbolic tangent function:

    \begin{equation}
        \begin{aligned}
            k(s) = - \frac{\tanh((s - L)/l)}{2 R} + \frac{\tanh((s - L - \pi R)/l)}{2 R} - \\ -  \frac{\tanh((s - 3 L - \pi R)/l)}{2 R}  + \frac{\tanh((s - 3 L - 2 \pi R)/l)}{2 R}
        \end{aligned}
    \end{equation}

\noindent By expressing the curvature of the condensate film surface in terms of the curvature of a flattened tube, we obtain the following equation:

    \begin{equation}
    \label{eq:k_film}
        k_f = \frac{(1 + kh)(k(1 + kh) - h'') + h'(2kh' + k'h)}{((1 + kh)^2 + (h')^2)^{3/2}}
    \end{equation}

\noindent and in the limit $hk \ll 1$, Eq. \eqref{eq:k_film} simplifies as:

    \begin{equation}
        k_f = \frac{k - h'' + 2k(h')^2 + k'hh')}{(1 + (h')^2)^{3/2}}
    \end{equation}

Since the curvature $k$ and height $h$ are functions of $s$, we can introduce the derivative of the liquid surface's curvature $k_f' = \frac{dk_f}{ds}$ as follows:

    \begin{eqnarray}
            \nonumber
            k_f' & = & \frac{k' -h''' + 3k'(h')^2 + 2k^2h' + 3kh'h'' + k''hh'}{(1 + (h')^2)^{3/2}}\\
            && - \frac{3h'h''(k - h'' + 2(h')^2 k + k'hh')}{(1 + (h')^2)^{5/2}}
    \end{eqnarray}

Assuming invariance by translation along the axis of the tube, and in the absence of shear stresses on the surface of the condensate film, the evolution equation for the liquid condensation on the inner surface of the tube, Eq. \eqref{nonlinear}, thus takes the simpler form:

    \begin{equation}
        h_f + \frac{\partial}{\partial s} \left[\frac{\rho h^3}{3 \mu} \left(-\sigma {k'}_{f}(s) + \rho g \sin\left(\alpha(s) - \arctan(h)\right)\right)\right] - \frac{\lambda \Delta T}{\rho \Psi h} = 0
    \end{equation}

\noindent which is the equation that we numerically solve. 

\section{Results and Discussion}
\label{sec4}

    \subsection{Dynamics of liquid-vapor motion}
    \label{subsec4.1}
				The flattened tubes are modeled as being a deformation of a circular tube, ensuring that the heat transfer surface area remain constant. Water is selected as the working fluid, with the wall temperature uniformly maintained. The gravitational field is assumed uniform with acceleration \(9.81 \, \text{N} \cdot \text{kg}^{-1}\).
        
\begin{figure}[hbt!]
            \centering
            \includegraphics[scale = 0.45]{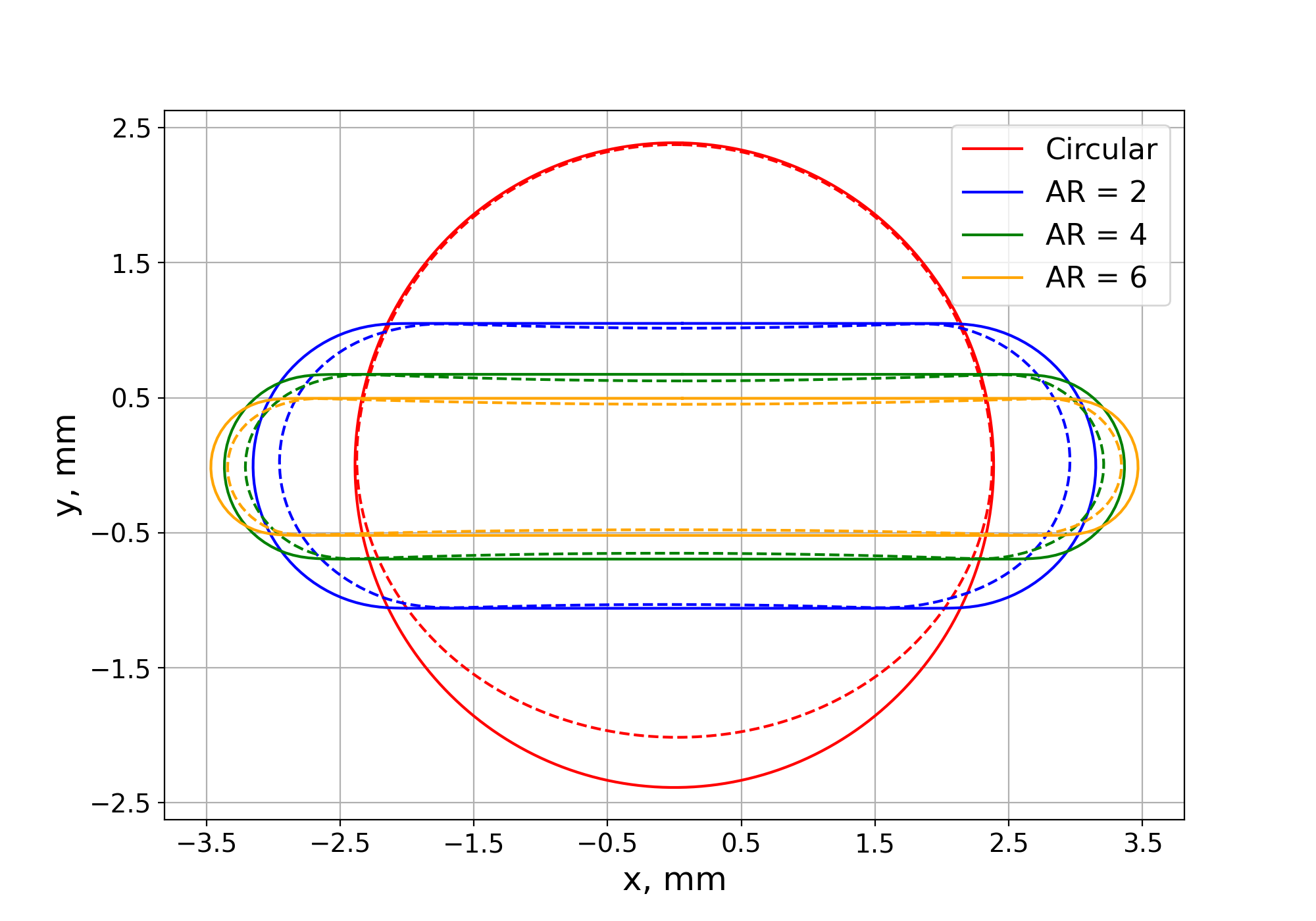}
            \caption{Profile cross-section with water/vapor interface for different aspect ratios for void fraction 0.9.}
            \label{fig:cross-section}
        \end{figure}

\begin{figure}[hbt!]
            \centering
            \includegraphics[scale = 0.45]{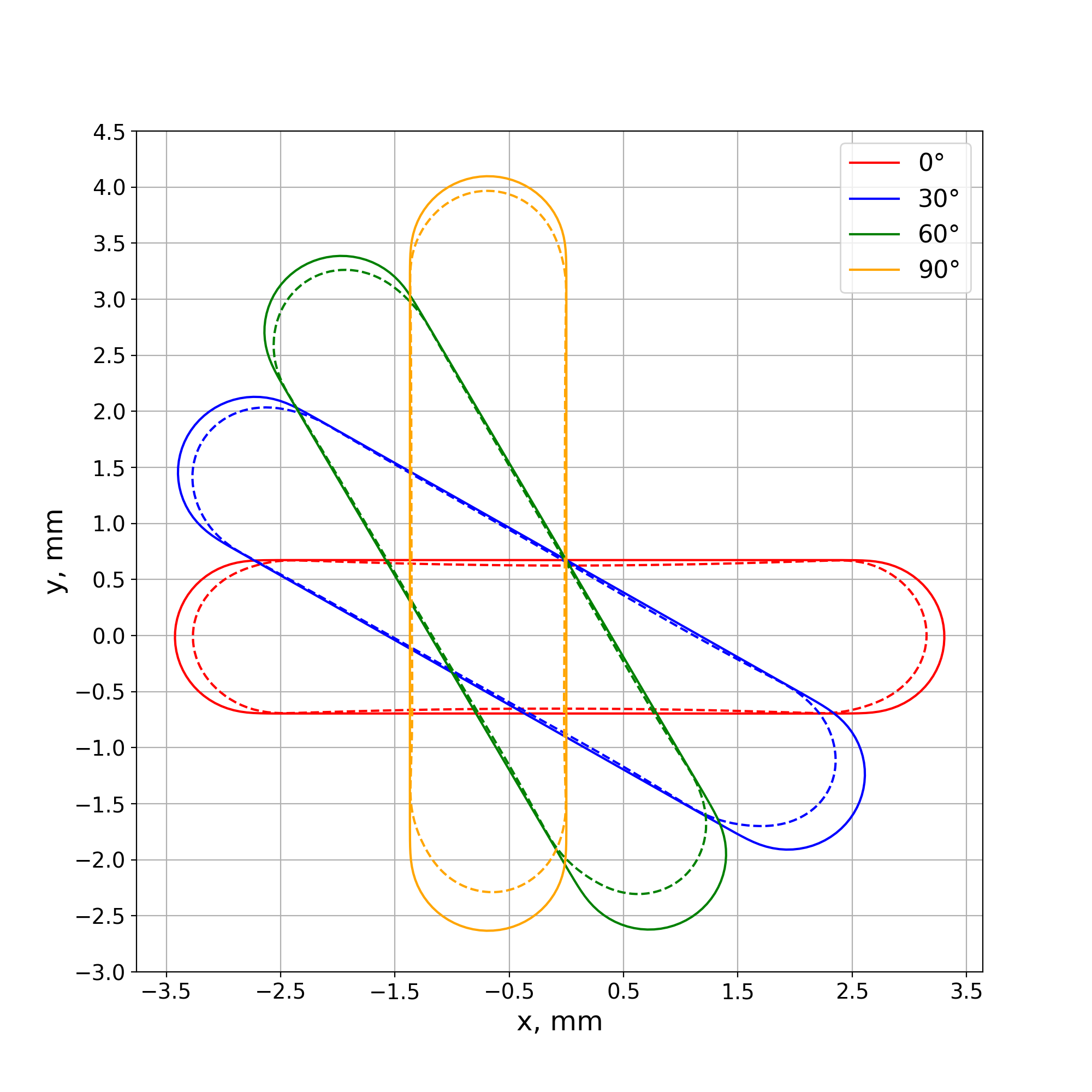}
            \caption{Profile cross-section with water/vapor interface for different rotation angles and for a void fraction equal to 0.9.}
            \label{fig:cross-section AR}
        \end{figure}

        The liquid film thickness, defined as the distance between the liquid–vapor interface and the tube wall, was calculated for different aspect ratios and profile rotation angles. The local and average heat transfer coefficients were found to be dependent on the film thickness and its distribution. The average heat transfer coefficient was determined for each scenario based on the distribution of the liquid layer thickness. The assessment of heat transfer improvement was based on the void fraction $\phi$, defined as the ratio of the vapor's surface to the total tube's surface:
        
        \begin{equation}
        \label{voidfraction}
            \phi = \frac{S_{\text{vapor}}}{S_{\text{vapor}} + S_{\text{liquid}}}
        \end{equation}

\noindent with $\phi$ varying from 0 (pure liquid) to 1 (pure vapor). Figures \ref{fig:cross-section} and \ref{fig:cross-section AR} illustrate the profile cross-sections with liquid-vapour interfaces for the same void fraction $(\phi \approx 0.9)$.
        
\begin{figure}[hbt!]
    \centering
    \includegraphics[width=1\linewidth]{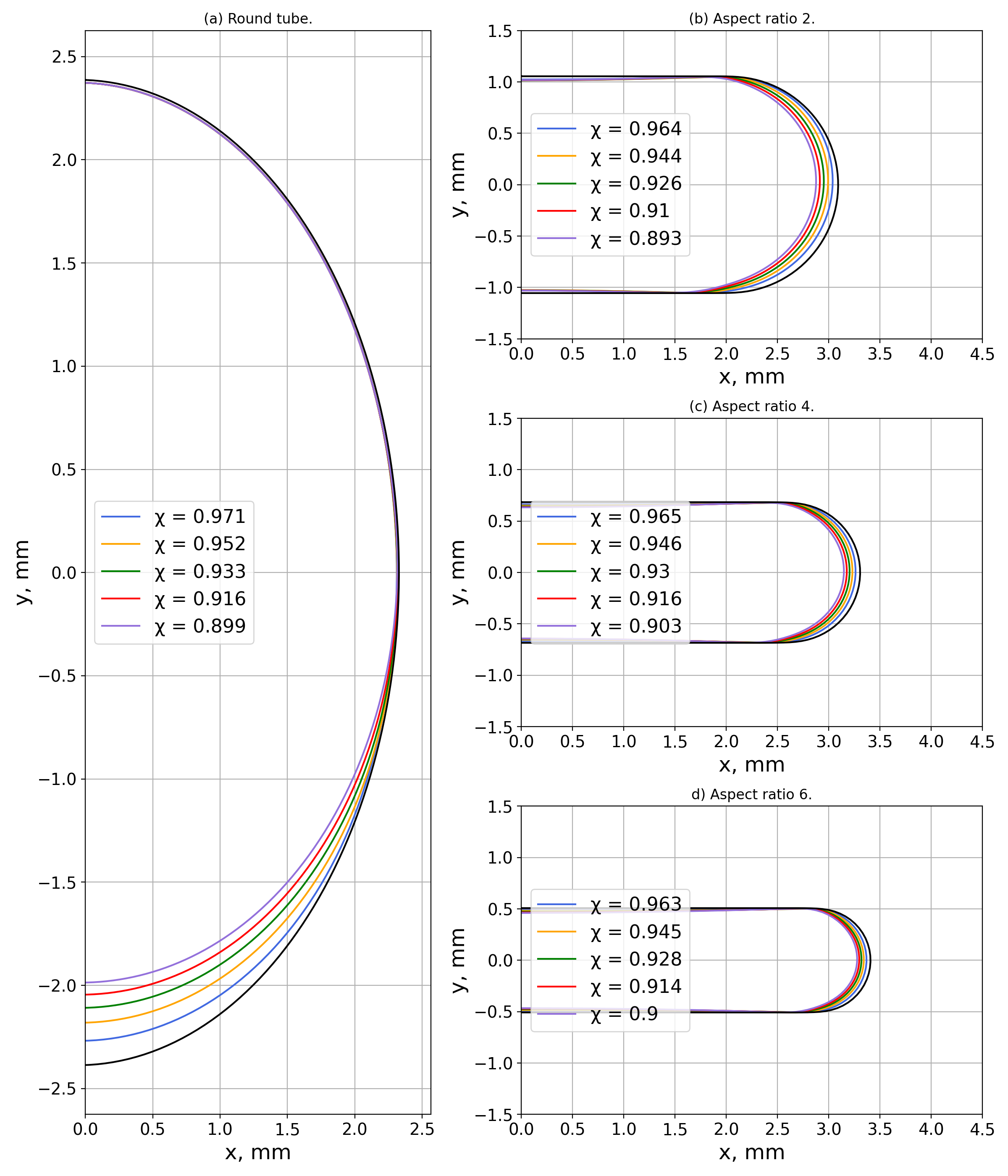}
    \caption{Water vapor interfaces in round tube and flattened tubes.}
    \label{fig:Vapors}
\end{figure}

To gain a better understanding of the condensation process in both circular and flattened tubes, the locations of liquid–vapor interfaces were identified. In circular tubes (Fig. \ref{fig:Vapors} (a)), the liquid film tends to accumulate at the bottom because of gravity. The accumulation becomes more pronounced with a decreasing void fraction. At higher void fraction levels, the liquid film further spreads over the tube surface, adopting an axisymmetric shape.

In flattened tubes (Figures \ref{fig:Vapors} (b), \ref{fig:Vapors} (c), and \ref{fig:Vapors} (d) for aspect ratios of 2, 4, and 6, respectively), the fluid accumulates in the rounded sections of the tubes, with a very thin liquid film forming on the flat sections. The thinnest liquid layer is at the transitions from rounded to flat sections. Note that capillary pressure gradients cause the liquid to move towards the rounded sections, leaving a thin film on the flat parts. This effect, known as the ``Grigorig'' effect \cite{stefano2012numerical}, becomes more pronounced with larger aspect ratios. The capillary pressure gradient is influenced by buoyancy effects due to gravitational forces and by the density difference between liquid and vapor, leading to a redistribution of the condensate layer up the tube walls.

\begin{figure}
    \centering
    \includegraphics[width=1\linewidth]{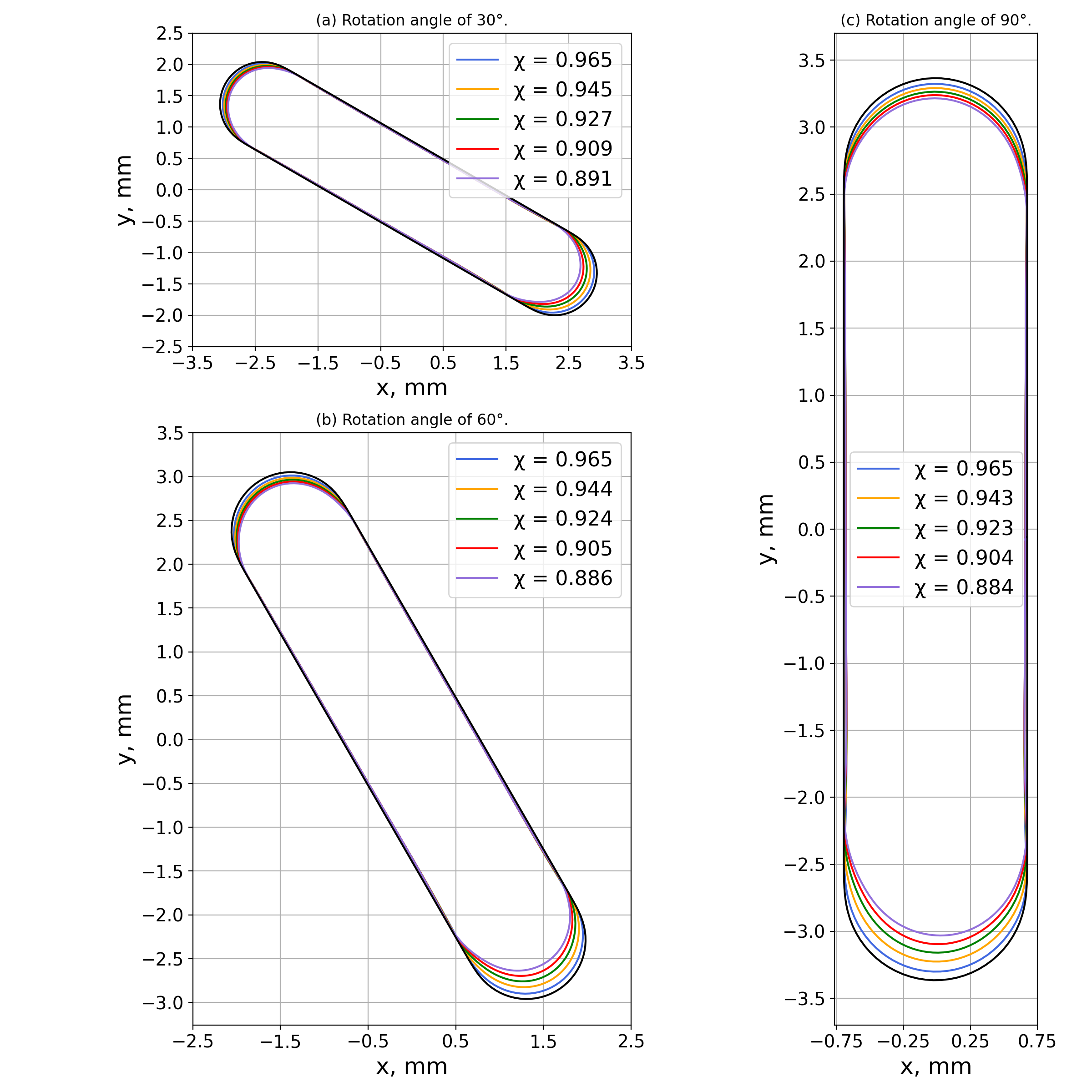}
    \caption{Water vapor interfaces in flattened tubes with aspect ratio 4 and different rotation angles.}
    \label{fig:Vapors AR4}
\end{figure}

Different rotation angles of the flattened tubes (Figures \ref{fig:Vapors AR4} (a), \ref{fig:Vapors AR4} (b), and \ref{fig:Vapors AR4} (c)) show a consistent trend: Regardless of the rotation angle, the liquid accumulated in the rounded sections. As the void fraction increases, the layer thickness predominantly increases in the rounded parts, partly because of surface tension and partly because of the influence of gravitation. The main difference observed with varying rotation angles is that the condensate layer in the lower rounded part is notably thicker than it is in the upper part, with this discrepancy growing as the rotation angle approaches 90$^\circ$.

Hence, the simulation results indicate that both aspect ratio and profile rotation angle significantly influence heat transfer in flattened tubes. Higher aspect ratios and increased rotation angles result in better heat transfer performance due to the redistribution of the condensate layer driven by capillary pressure gradients and gravitational forces.

\subsection{Film thickness distribution throughout the flattened tubes for various aspect ratios}
\label{subsec4.2}

    \begin{figure}
        \centering
        \includegraphics[width=1\linewidth]{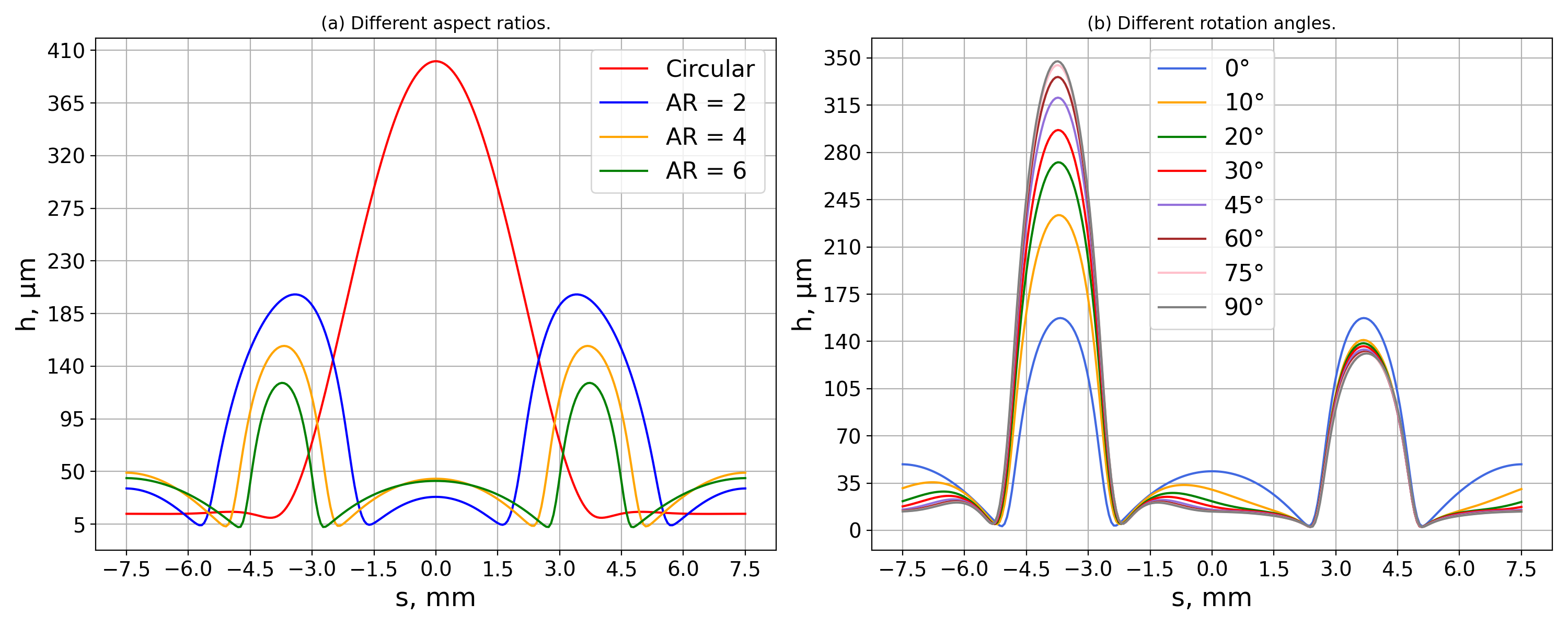}
        \caption{Film thickness distribution along the perimeter of a flattened tube for a void fraction of 0.9 and different aspect ratios (a) and rotation angles (b).}
        \label{fig:Film thickness distribution all}
    \end{figure}

The average film thickness, $\bar{h}$, is calculated as the integral mean value along the perimeter of the flattened tube's inner surface section:
    
    \begin{equation}
    \label{average film thickness}
        \bar{h} = \frac{1}{s} \int_{0}^{s} h(s) \, ds    
    \end{equation}
		
\noindent Figure \ref{fig:Film thickness distribution all} (a) provides a clear depiction of how the condensate film redistributes in the tubes, thus enabling an easy estimation of the local layer thicknesses. Notably, the film thickness minimum is observed at four distinct points, specifically at the junctions between the rounded and flat sections of the tube. We observe that as the aspect ratio increases, these points of minimum thickness appear lower on the graph, indicating more pronounced local minima. Conversely, the maxima of the thickness curve, which correspond to the rounded sections of the tubes, show that the greatest local layer thickness occurs for the smallest aspect ratio for a given void fraction. This thickness decreases as the aspect ratio increases. As regards the flat sections, the tube with the smallest aspect ratio exhibits the thinnest condensate layer at the lower part of the tube. However, in the upper flat area, the results are more complex due to the varying balance between surface tension forces and gravitational forces. 

Figure \ref{fig:Film thickness distribution all} (b) provides additional insights on the condensate layer distribution at various rotation angles. The curves were generated under consistent conditions: a single aspect ratio $(AR = 4)$ and void fraction $(\phi \approx 0.9)$. The coincidence of all local minima confirms that the aspect ratio primarily dictates the minimum film thickness. In the rounded sections of the flattened tubes, the thickness of the condensate film increases, with the maximum thickness observed in the lower rounded section due to the combined effects of gravity and surface tension. The prominence of this maximum increases with the angle of rotation. However, as the rotation angle increases, the thickness in the upper rounded section decreases. Moreover, an increase in the rotation angle causes the film thickness on the flat parts of the tube to decrease, shifting the distribution towards the lower part of the tube due to gravity.

\begin{figure}[hbt!]
        \centering
        \includegraphics[scale = 0.45]{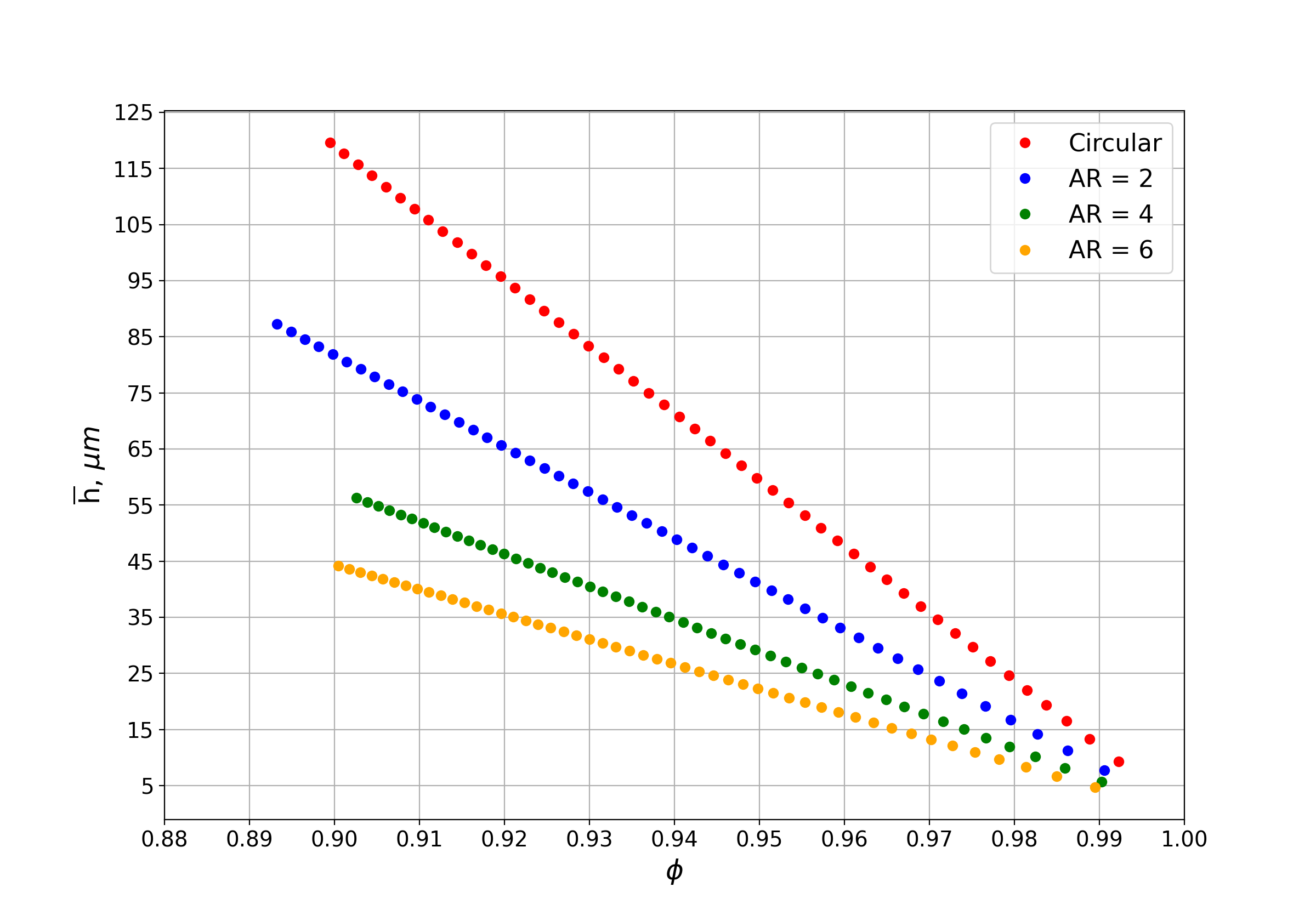}
        \caption{Average film thickness dependence on the void fraction for different aspect ratios.}
        \label{fig:average film thickness}
    \end{figure}

As depicted in Figure \ref{fig:average film thickness}, the average film thickness decreases with increasing aspect ratio. Additionally, the average film thickness in flattened tubes is significantly smaller compared to that in a circular tube. This reduction in film thickness is attributed to increased shear stress.
    
The present analysis thus demonstrates that increasing the aspect ratio of flattened tubes leads to a more pronounced local minimum of the condensate film thickness, thereby enhancing heat transfer efficiency. The distribution and thickness of the condensate film are significantly influenced by both the aspect ratio and the rotation angle of the tube profile, with gravity and surface tension playing pivotal roles in their dynamics.

\subsection{Distribution and improvement of the heat transfer coefficient for various aspect ratios and rotation angles}
\label{subsubsec4.3}

\begin{figure}[hbt!]
            \centering
            \includegraphics[width=1\linewidth]{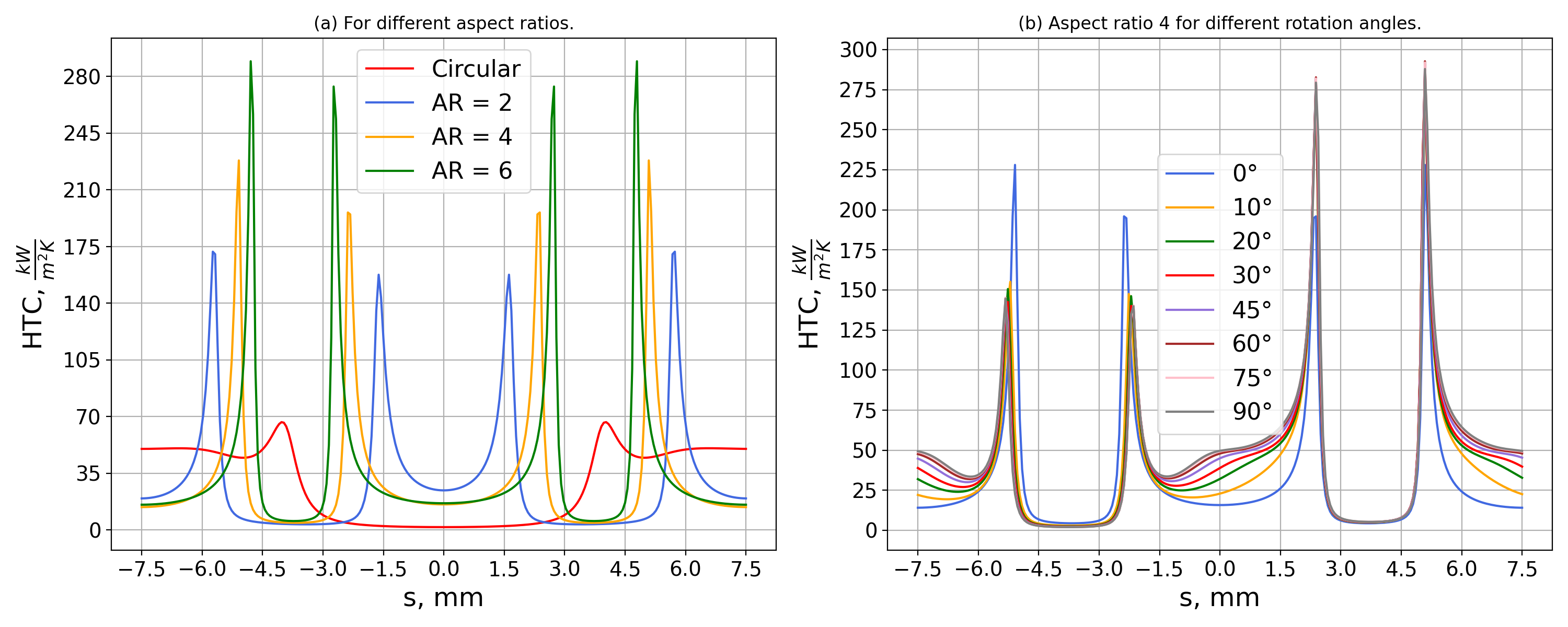}
            \caption{HTC distribution along the perimeter of a flattened tube with void fraction 0.9.}
            \label{fig:HTC distribution}
        \end{figure}

        The heat transfer coefficient is inversely related to the film thickness: HTC $= \lambda/h$, with $\lambda$ being the thermal conductivity. This relationship implies that a thicker condensate layer results in a lower heat transfer coefficient. The distributions of HTC for various configurations of flattened tubes are depicted in Figures \ref{fig:HTC distribution} (a) and \ref{fig:HTC distribution} (b), which show the variations for different aspect ratios and rotation angles, respectively. Notably, the maxima of HTC occur at the junctions between the flat and rounded sections of the tubes.

To evaluate the heat transfer enhancement, it is essential to calculate the average HTC along the perimeter of the flattened tubes: 
        
        \begin{equation}
        \label{HTC average}
            \overline{\text{HTC}} = \frac{1}{s} \int_{0}^{s} \text{HTC}(s) \, ds
        \end{equation}

\noindent This integral calculation provides an informative measure of the heat transfer performance across different aspect ratios and rotation angles. Using the mathematical model of thin films described in Section \ref{sec3}, we assess the impact of varying aspect ratios on the heat transfer process in flattened tubes. AR is ranging from 2 to 6, which are the commonly used aspects ratios \cite{stefano2010numerical, jian2018numerical, kim2013condensation, lee2014condensation}. Note that we our numerical model, we can effectively simulate the growth of a thin condensate layer, though it has limitations when the layer thickness increases, which restricts the void fraction range. Despite these limitations, our model accurately reflects the actual condensation processes in flattened tubes used in heat exchangers.

\begin{figure}[hbt!]
            \centering
            \includegraphics[scale = 0.45]{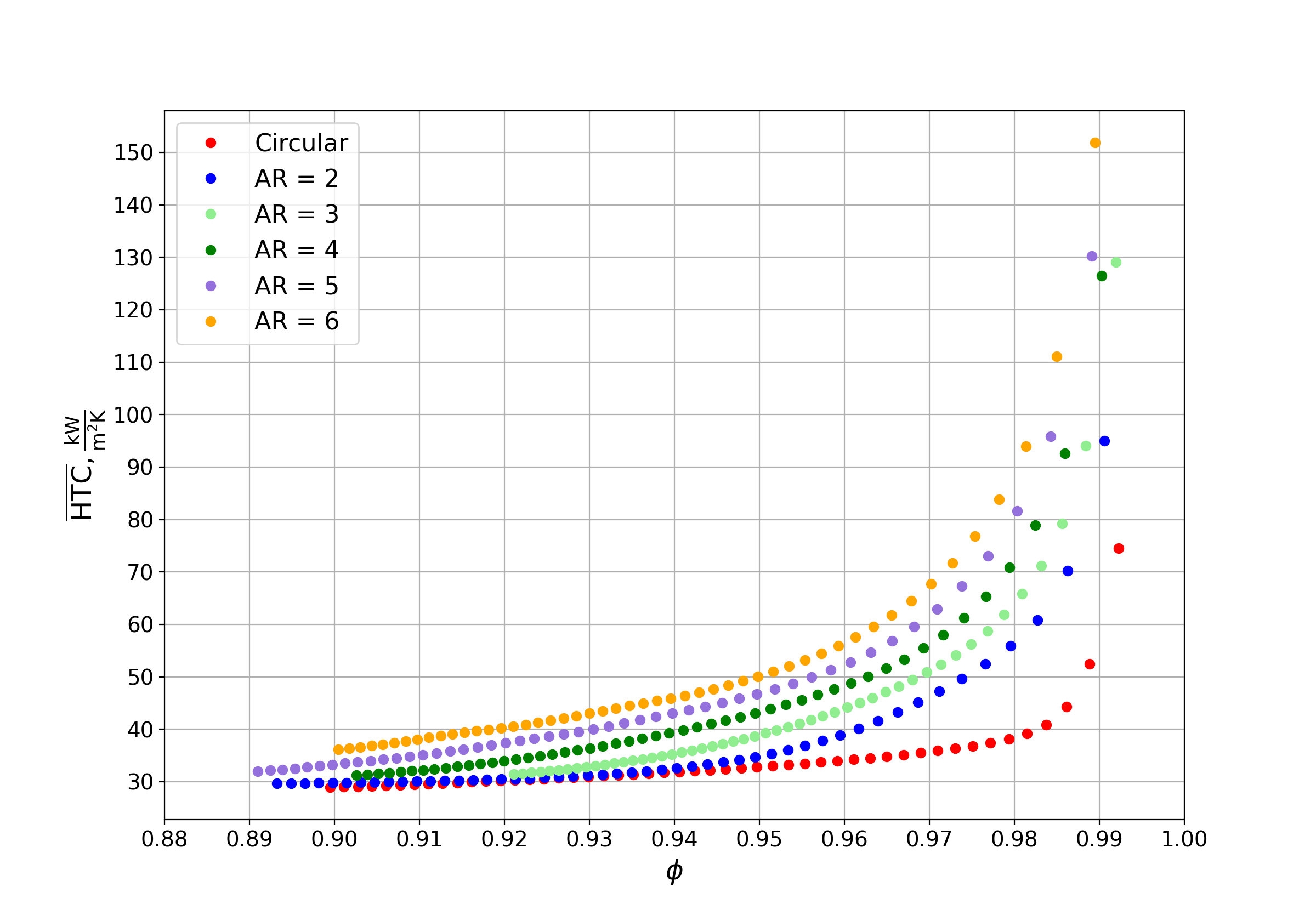}
            \caption{Average heat transfer coefficients for flattened tubes of different aspect ratios comparing with the circular tube for void fraction 0.9.}
            \label{fig:Average heat comparing AR CT}
        \end{figure}

Figure \ref{fig:Average heat comparing AR CT} shows that the average HTC increases with both aspect ratio and void fraction. Flattened tubes thus significantly enhance heat transfer compared to circular tubes with the same perimeter. This observation is consistent with previous experimental and numerical studies by Kim et al. \cite{kim2013condensation}, Nebuloni and Thome \cite{stefano2012numerical}, and Wen et al. \cite{jian2018numerical}. 
        
\begin{figure}[hbt!]
            \centering
            \includegraphics[scale = 0.45]{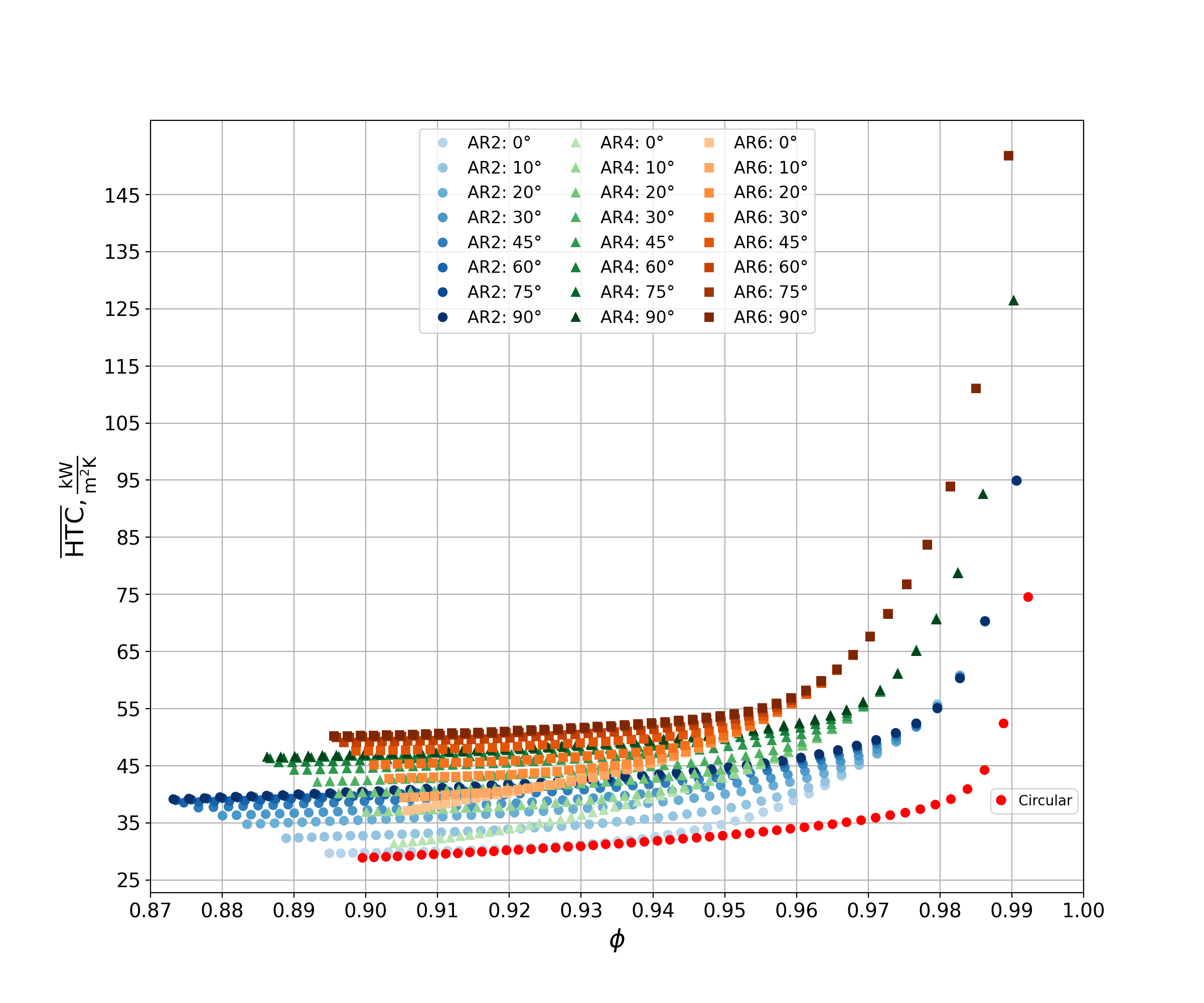}
            \caption{Average heat transfer coefficients for flattened tubes of different aspect ratios and rotation angles compared with the circular tube for void fraction 0.9.}
            \label{fig:Average heat comparing AR RA CT}
        \end{figure}

In the early stages of condensate layer formation, the liquid is evenly distributed over the flattened tubes due to the negligible effect of gravity on the small amount of fluid. Indeed, surface tension forces are the main cause of the differences observed among the distributions of the average heat transfer coefficient for varying aspect ratios as shown in Figure \ref{fig:Average heat comparing AR RA CT}. The differences in average HTC for various angles increase with the condensate film thickness and decreasing void fraction. A higher rotation angle results in a greater average HTC, primarily due to the accumulation of liquid in the lower rounded sections of the flattened tubes.
        
\begin{figure}[hbt!]
            \centering
            \includegraphics[scale = 0.4]{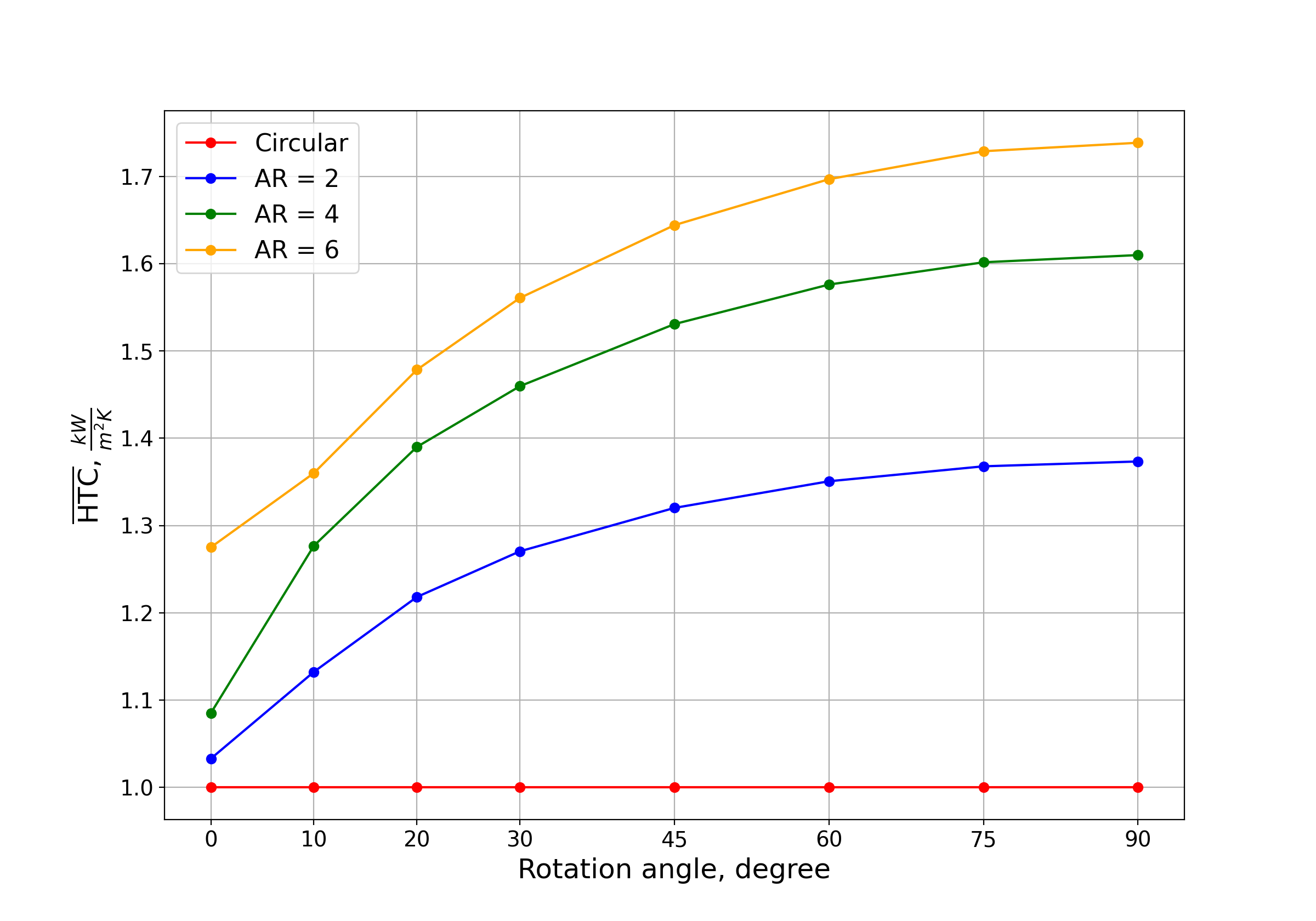}
            \caption{Average heat transfer coefficients for flattened tubes of different aspect ratios and rotation angles compared with the circular tube for void fraction 0.9.}
            \label{fig:void fraction 0.9}
        \end{figure}

\begin{figure}[hbt!]
            \centering
            \includegraphics[scale = 0.4]{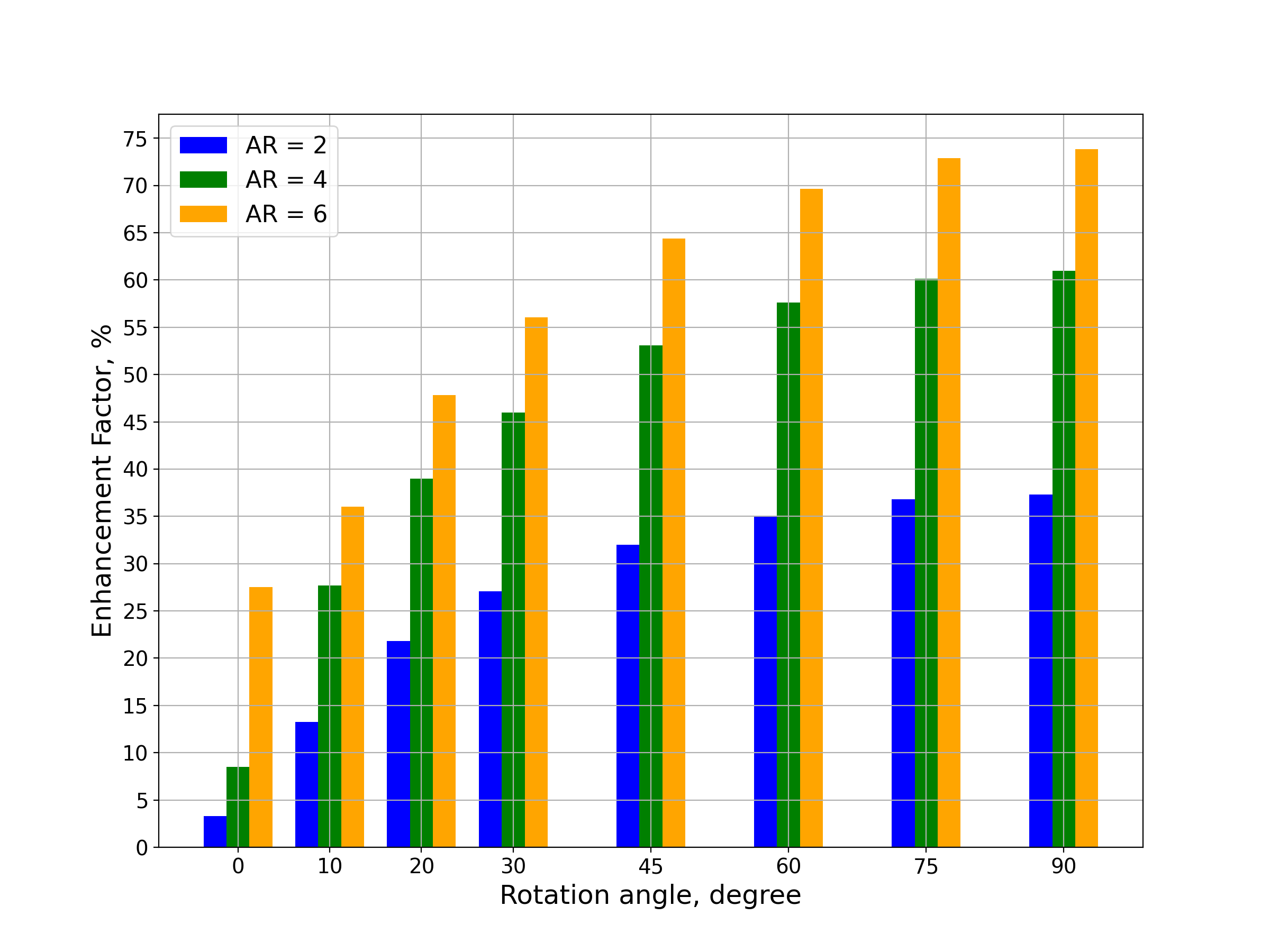}
            \caption{Enhancement factor for flattened tubes of different aspect ratios and rotation angles for void fraction 0.9.}
            \label{fig:Enhancement factor}
        \end{figure}

Figures \ref{fig:void fraction 0.9}, \ref{fig:Enhancement factor} present a specific analysis for a void fraction of 0.9, focusing on the impact of rotation angles. Increasing both the aspect ratio and the rotation angle significantly enhances the HTC compared to a circular tube. The enhancement factor (EF), defined as the percentage increase in the average HTC of modified flattened tubes relative to a circular tube with the same heat transfer surface:

\begin{equation}
           \text{EF} = \left( \frac{\overline{\text{HTC}}_{\text{AR}+\text{rotation angles}}}{\overline{\text{HTC}}_{\text{round tube}}} - 1 \right) \times 100\%
\end{equation}

\noindent is shown in Figure \ref{fig:void fraction 0.9}. At a zero rotation angle, the EF for aspect ratios of 2, 4, and 6 are 3\%, 9\%, and 28\%, respectively. The difference between aspect ratios 2 and 6 across various angles is approximately 30\%, reaching up to 37\% at a 90$^\circ$ rotation angle. The maximum EF of 74\% is achieved with an aspect ratio of 6 and a 90$^\circ$ rotation angle, driven by the capillary pressure gradient. This gradient causes most of the liquid to accumulate in the small rounded section of the tube.        

So, increasing the aspect ratio and rotation angle of flattened tubes significantly improves heat transfer performance due to the optimized distribution of the condensate film. These findings provide valuable insights for the design and optimization of heat exchangers using flattened tube profiles.

\section{Conclusion}
\label{sec5}
We developed a numerical model for the prediction of the laminar film condensation of a pure vapour inside a flattened tube. Anumerical study of the condensation process was performed for horizontal circular and flattened tubes with water as a working substance. The effects of the flattened tube aspect ratio and rotation angle on heat transfer characteristics were investigated. Circle tubes with a perimeter of 15 mm were deformed into flattened tubes with aspect ratios ranging from 2 to 6. The rotation angle varied from 0 to 90 degrees. The thickness of the condensed liquid film and the void fraction were analyzed in detail in order to understand the interaction between capillary and gravitational forces, as well as the patterns of the vapour-liquid interface and heat transfer during condensation inside the flattened tubes.
    
The numerical simulations clearly show that the heat transfer coefficients of flattened tubes increase with the aspect ratio (3\% - AR2, 9\% - AR4, and 28\% - AR6). The capillary and gravity forces play a key role in forming the vapour-liquid interface that determines the local thermal performance inside a flattened tube. The main volume of the liquid is collected in arc sections of the tube because of surface tension. It leads to thinning the liquid thickness in straight tube sections where higher heat transfer coefficients are achieved. The heat transfer enhancement increases with aspect ratio and rotation angle (from 3\% to 37\% - AR2, from 9\% to 61\% - AR4 and from 28\% to 74\% - AR6) as shown in Fig.~\ref{fig:Enhancement factor}. For an aspect ratio of 6 and a rotation angle of $90^{\circ}$, the maximum EF of $74\%$ is attained. We hope that this level of performance enhancement found numerically will stimulate experimental research for testing our results and further extend the scope of the analysis presented here. 

\bibliographystyle{elsarticle-num}
\bibliography{main}

\end{document}